\documentclass[useASM, referee]{mn2e}
\usepackage{graphicx}
\usepackage{lscape}
\usepackage{amssymb}
\usepackage{setspace}

\begin{document}
%\linenumbers
\singlespacing

\title[Newborn magnetars in GRBs]
{Comparison of the Characteristics of Magnetars Born in Death of Massive Stars and Merger of Compact Objects With {\em Swift} Gamma-Ray Burst Data}

\author[Zou et al.]
{Le Zou$^{1}$, En-Wei Liang$^{1}$\thanks{E-mail:
lew@gxu.edu.cn}, Shu-Qing Zhong$^{2}$, Xing Yang$^{1}$, Tian-Ci Zheng$^{1}$, Ji-Gui Cheng$^{1}$,
\newauthor{Can-Min Deng$^{1}$, Hou-Jun L\"{u}$^{1}$, Shan-Qin Wang$^{1}$} \\
$^1$Guangxi Key Laboratory for Relativistic Astrophysics, School of Physical Science and Technology, Guangxi University, Nanning 530004, China\\
$^2$Department of Astronomy, School of Physical Sciences, University of Science and Technology of China, Hefei 230026, China\\}
\maketitle

\label{firstpage}
\begin{abstract}
Assuming that the shallow-decaying phase in the early X-ray lightcurves of gamma-ray bursts (GRBs) is attributed to the dipole radiations (DRs) of a newborn magnetar, we present a comparative analysis for the magnetars born in death of massive stars and merger of compact binaries with long and short GRB (lGRB and sGRB) data observed with the {\em Swift} mission. We show that the typical braking index ($n$) of the magnetars is $\sim 3$ in the sGRB sample, and it is $\sim 4$ for the magnetars in the lGRB sample. Selecting a sub-sample of the magnetars whose spin-down is dominated by DRs ($n\lesssim 3$) and adopting a universal radiation efficiency of $0.3$, we find that the typical magnetic field strength ($B_p$) is $10^{16}$ G {\em vs.} $10^{15}$ G and the typical initial period ($P_0$) is $\sim 20$ ms {\em vs.} $2$ ms for the magnetars in the sGRBs {\em vs.} lGRBs. They follow the same relation between $P_0$ and the isotropic GRB energy as $ P_0\propto E_{\rm jet}^{-0.4}$. We also extend our comparison analysis to superluminous supernovae (SLSNe) and stable pulsars. Our results show that a magnetar born in merger of compact stars tends to have a stronger $B_p$ and a longer $P_0$ by about one order of magnitude than that born in collapse of massive stars. Its spin-down is dominated by the magnetic DRs as old pulsars, being due to its strong magnetic field strength, whereas the early spin-down of magnetars born in massive star collapse is governed by both the DRs and gravitational wave (GW) emission. A magnetar with a faster rotation speed should power a more energetic jet, being independent of its formation approach.
\end{abstract}

\section {Introduction}
The remnant of collapse of massive stars or merger of compact stars may be a magnetar, which can serve as the central engine of some gamma-ray bursts (GRBs; e.g. Usov 1992; Thompson 1994; Dai \& Lu 1998a, 1998b; Zhang \& M\'{e}sz\'{a}ros 2001; Metzger et al. 2008, 2011; see review by Woosley \& Bloom 2006; Kumar \& Zhang 2015). This speculation is convinced by observations of a shallow decay segment in the early X-ray lightcurves of a large fraction of GRBs detected by the X-Ray Telescope (XRT) onboard the {\em Swift} mission (e.g. Nousek et al. 2006; Zhang et al. 2006; Liang et al. 2007; Metzger et al. 2008). This segment is consistent with the temporal evolution feature of the kinetic luminosity injected from dipole radiations (DRs) of a magnetar before its characteristic spin-down timescale ($\tau$). It was also proposed that the soft, steady extended gamma-rays observed in some GRBs are from the magnetar DRs (Barthelmy et al. 2005; Gehrels et al. 2006; Lin et al. 2008; Perley et al. 2009; Zhang et al. 2009; Bucciantini et al. 2012; Gompertz et al. 2013), and late X-ray flares are from the late re-activities of a magnetar central engine (Dai et al. 2006). Particularly, the shallow-decay segment in the X-ray lightcurves of some GRBs, such as GRB 070110, features as a plateau with a sharp drop, indicating that the X-rays would be from an internal energy dissipation process and the sharp flux drop is due to the cease of the process (Troja et al. 2007). Such an internal plateau is recognized as ``Smoking Gun'' evidence for collapse of a supra massive magnetar into a black hole (Troja et al. 2007; Liang et al. 2007; Lyons et al. 2010; Yu et al. 2010; L\"{u} \& Zhang 2014; see also review by Kumar \& Zhang 2015). The ``Smoking Gun" signature is also observed in the X-ray lightcurves of some sGRBs. Rowlinson et al. (2013) showed that about half of the lightcurves of sGRBs observed with the {\em Swift}/XRT from Dec. 2004 to May 2012 can be fitted with the magnetar DR model.

The magnetic dipole radiation would be a probe for the properties of a newborn magnetar, such as its magnetic field strength ($B_p$), initial period ($P_0$), braking index ($n$), and equation of state (EoS). Lyons et al. (2010) derived $B_p$ and $P_0$ values with a sample of long GRBs (lGRBs) observed by the {\em Swift} mission (see also L\"{u} \& Zhang 2014). L\"{u} et al. (2019) found that the braking indices of newborn magnetars in lGRBs are $2\sim 5$, likely hinting that the gravitational wave (GW) emission may dominate the spin-down of some newborn magnetars in their lGRBs sample. It was also found that a magnetar with a faster-rotating speed may power a more energetic GRB (Zou et al. 2019). The hydrodynamic behavior of a magnetar produced from mergers of a neutron star (NS) binary depends on its EoS and the total mass of the NSs (Hotokezaka et al. 2013). L\"{u} et al. (2015) derived the parameters of the magnetars in some sGRBs and constrained the EoS of the magnetars in these sGRBs (seel also Gao et al. 2016; Piro et al. 2017).

It is interesting whether the characteristics of magnetars born in collapse of massive stars and compact binary mergers have any difference. In addition, a magnetar may also be the energy source of some superluminous supernovae (SLSNe). Are the newborn magnetars in both sGRBs and lGRBs different from those in SLSNe? This paper aims to explore these questions. Our sample and data analysis are presented in \S 2. We derive the physical parameters of the magnetars and analyze their possible correlations in \S 3. Discussions on our analysis results are presented in \S 4. Summary and conclusion are presented in \S 5. Throughout, a concordance cosmology with parameters $H_0 = 71$ km s$^{-1}$ Mpc $^{-1}$, $\Omega_M=0.30$, and $\Omega_{\Lambda}=0.70$ is adopted.

\section{Sample and Data Analysis}
Our GRBs samples are selected from the current {\em Swift} GRB catalog. Their prompt gamma-ray lightcurves observed with the Burst Alert Telescope (BAT) and X-ray afterglow lightcurves observed with XRT are taken from the {\em Swift} data archive (Evans et al. 2010). The lGRB sample is taken from Zou et al. (2019). It includes 67 bursts with redshift measurement. For the sGRBs sample, we adopt the same criterion as that for the lGRBs. We fit their XRT lightcurves post the early steep decay segment or X-ray flares with a smooth broken power-law, $F=F_0[(t/t_b)^{\alpha_1}+(t/t_b)^{\alpha_2}]^{-1}$, and select those GRBs whose early XRT lightcurves are composed of a shallow-decay segment, which is defined as $\alpha_1<0.75$ (e.g. Liang et al. 2007, 2008). Our final sGRB sample includes 28 sGRBs. The derived slopes ($\alpha_1$ and $\alpha_2$), break time ($t_b$) of their lightcurves and the corresponding luminosity at the $t_b$ are reported in Table 1. Among the 28 sGRBs, 15 bursts have redshift ($z$) measurement. GRB 060614 is also included in our sGRB sample since it may be originated from a binary merger as proposed by some groups (Gehrels et al. 2006; Gal-Yam et al. 2006; Della Valle et al. 2006; Fynbo et al. 2006; Zhang et al. 2007; Yang et al. 2015).

The Pearson correlation analysis is adopted to evaluate the possible linear correlation among variables. The Pearson correlation coefficient $r$ refers to the degree of a linear correlation between two variables, i.e. a larger $|r|$ value indicates a stronger correlation, and the $p$-value gives the chance probability of such a correlation, which is statistically acceptable in case of $p<10^{-4}$. Note that the data uncertainties are not taken into account in the Pearson correlation analysis. We employ the Monte Carlo simulation to evaluate the influence of the data uncertainties on the correlation analysis. Assuming that the observational errors are normally distributed, we bootstrap $10^4$ data sets and calculate the $r$ and $p$ values for each data set, and derive their central values and $1\sigma$ confidence level from the $r$ and $p$ distributions. The least square fit method is used for deriving a relation between two variables by considering their uncertainties. We also adopt the adjusted $R^2$ ($R^2_a$) to measure the fitting results. A $R^2_a$ value closer to 1 indicates a better fit. The statistical difference between two data sets is evaluated with the probability $p_{\rm KS}$ of the Kolmogorov-Smirnov test (K-S test). In the case of $p_{\rm KS}<10^{-4}$, the two data sets are statistically claimed from different parents.

We extract the spectra of the prompt gamma-ray emission and the early X-ray flares. The total fluence of a GRB jet is the sum of observed fluences in the BAT-XRT band. Note that the GRB spectra in the keV-MeV band are usually fitted with a smooth broken power-law function, the so-called Band function (Band et al. 1993). BAT is sensitive in the energy range of 15-150 keV. The GRB spectra observed with BAT are usually adequate fit with a single power-law function (e.g. Zhang et al. 2006). This may lead to a systematical underestimate of the GRB fluences. {\em Konus}/Wind is sensitive in a much wider energy band (~20keV-2MeV) than BAT. It may collect a relatively hard GRB sample. We adopt a sample of 168 GRBs that were simultaneously detected by BAT and {\em Konus}/Wind from Tsvetkova et al. (2021) to evaluate the systematical impact of the single power-law function on our estimate of the burst fluences. As shown in Figure 1, the fluences observed by the two detectors are corrected. Our Pearson correlation analysis gives a coefficient of $0.87\pm 0.01$ and a chance probability of $p\sim1.23\times10^{-51}$. The fluences observed with {\em Konus}/Wind are systematically larger than the BAT fluences, typically with a factor of $\sim 2.5$. Therefore, this effect does not significantly affect our estimate of the fluences for the sGRB in our sample. We still adopt the single power-law model ($N(E)\propto E^{-\Gamma_{\rm jet}}$) to derive the gamma-ray fluences observed with BAT for our analysis. The spectra of the X-ray flares are fitted with an absorbed single power-law model. The isotropic jet energy is calculated by $E_{\rm jet}=4\pi D^2_L (S_\gamma+S_X)/(1+z)$, where $S_\gamma$ and $S_X$ are the prompt gamma-ray and X-ray flare fluences, $D_L$ is the GRB luminosity distance.
The energy of the DR wind is estimated with the X-ray data of the shallow decaying segment. We extract the spectrum of this segment and fit it with the absorbed single power-law model. The isotropic X-ray energetic is calculated with $E_{\rm wind}=4\pi D^2_L S_{\rm wind}/(1+z)$, where $F_{\rm wind}$ is the average flux of the shallow-decaying segment and $S_{\rm wind}$ is the observed X-ray fluence of the DR wind, i.e. $S_{\rm wind}=F_{\rm wind} t_b$.
The luminosity at $t_b$ is taken as the characteristic luminosity of the wind ($L_b$). Our results are reported in Table 1. Hereafter, we denote the data of the sGRBs and lGRBs with a superscript $s$ and $l$, respectively.

Figure \ref{tb-Lb} shows the $\log t_b$ and $\log L_{b}$ distributions along with $\log L_{b}$ as a function of $\log t_b$ in the burst frame. Measuring the difference of the $\log t_b$ distributions between the sGRB and the lGRB sample with the K-S test, we obtain $p_{\rm KS}<10^{-4}$, which indicates that the two distributions are statistically different. The $\log t^{l}_b$ distribution is log-normal. Our Gaussian fit gives $\log t^{l}_b/{\rm s}=4.11\pm 0.64$ ($1\sigma$), and the $R_a^2$ of the fit is 0.98. A tentative bimodal feature with a separation of $t_b=10^3$ seconds is observed in the $\log t^{s}_b$ distribution. The long-$t^{s}_b$ component is roughly consistent with that of the lGRBs. The short-$t^{s}_b$ component with a sharp drop at $10^{2}$ seconds. The sharp drop may be due to the observational bias for the restriction of XRT slewing to the bursts. It is also possible due to collapses of a newborn magnetar into a black hole before its characteristic spin-down timescale.
Among 15 sGRBs in the short-$t^{s}_b$ components, 8 sGRBs have an X-ray flux decay slope post $t^{s}_b$ being steeper than 3. Thus, the X-ray emission should be produced by an internal energy dissipation process. The sharp flux drop suggests the cease of this process, and the residual flux is due to the curvature effect for the photons from the high latitude. The most favorable scenario is that the newborn magnetar collapsed into a black hole before the spin-down characteristic timescale (e.g. Troja et al. 2007; Kumar \& Zhang 2015; L\"{u} et al. 2015). Therefore, the newborn magnetars in the 8 sGRBs may be in this scenario.

The $p_{\rm KS}$ of the $\log L^{l}_{b}$ and $\log L^{s}_{b}$ distributions are smaller than $10^{-4}$, suggesting that they are statistically different. The $\log L^{l}_{b}$ distribution can be fitted with a Gaussian function, yielding $\log L^{l}_{b}/{\rm erg s^{-1}}=48.27\pm 0.93$ and $R_a^{2}=0.94$. The $L^{s}_{b}$ distribution ranges from $2\times10^{44}$ erg s$^{-1}$ to $2\times10^{49}$ erg s$^{-1}$. The sGRBs and lGRBs form two marginally separated parallel sequences in the $\log L_b-\log t_{\rm b,z}$ plane, as shown in Figure \ref{tb-Lb}(c). Our best least square fit by the observational errors yields $\log L^{l}_b/{\rm erg\ s^{-1}}=(53.29\pm 0.48)-(1.60\pm 0.13)\log t_{\rm b,z}^{l}/{\rm s}$ and $\log L^{s}_b/{\rm erg\ s^{-1}}=(50.46\pm 0.85)-(1.25\pm 0.28)\log t_{\rm b,z}^{s}/{\rm s}$ with Pearson correlation coefficients of $-0.81\pm0.02$ and $-0.79 \pm 0.04$, respectively. The corresponding chance probabilities of these correlations are $p\sim 2.98\times10^{-16}$ and $p \sim 7.98\times 10^{-4}$. The radiation luminosity of the DR wind in the sGRBs is lower than that in the lGRBs by about three orders of magnitude, and its decay with time is shallower than that in the lGRBs.

Figure \ref{E-E} (a) shows the $\log E^{s}_{\rm jet}$ and $\log E^{l}_{\rm jet}$ distributions. They can be fitted with a Gaussian function, i.e. $\log E^{s}_{\rm jet}/{\rm erg}=50.66\pm 0.22$ (adjusted $R_a^{2}=0.90$) and $\log E^{l}_{\rm jet}/{\rm erg}=52.51\pm 0.38$ ($R_a^{2}=0.97$). The mean value of the $\log E^{s}_{\rm jet}$ distribution is about two orders of magnitudes lower than that of the lGRBs. Similarly, the $\log E^{l}_{\rm wind}$ distribution also can be fitted with a Gaussian function, i.e. $\log E^{l}_{\rm wind}/{\rm erg}=52.51\pm 0.38$, but the $E^{s}_{\rm wind}$ distribution is scattered in the range of $10^{48}\sim 10^{51}$ ergs, as shown in Figure \ref{E-E}(b). A very weak correlation between $S_{\rm wind}$ and $S_{\rm jet}$ is found with a Pearson correlation coefficient of $r=0.56\pm0.02$ and a chance probability of $p\sim2.2\times 10^{-8}$ for both the sGRBs and lGRBs, but no such a correlation can be statistically claimed for the sGRB sample only.

Figure \ref{gamma} shows the distributions of the spectral photon indices. It is found that the $\Gamma^{s}_{\rm jet}$ distribution is broader than that of the lGRBs. The $\Gamma^{s}_{\rm jet}$ for $40\%$ sGRBs in our sample is harder than 1, while the $\Gamma^{l}_{\rm jet}$ distribution ranges in $1\sim 2.5$, which can be fitted with a Gaussian function $\log \Gamma^{l}_{\rm jet}=1.77\pm 0.37$ ($1\sigma$) with an adjusted $R_a^2$ value as 0.94. The $\Gamma^{s}_{\rm wind}$ and $\Gamma^{l}_{\rm wind}$ distributions can be fitted with a Gaussian function, i.e. $\Gamma^{s}_{\rm wind}=1.73\pm 0.22$ and $\Gamma^{l}_{\rm wind}=1.96\pm 0.23$ with an $R_a^2$ value as 0.99 and 0.86, respectively. Although $\Gamma^{s}_{\rm wind}$ are averagely smaller than $\Gamma^{l}_{\rm wind}$, their means are still consistent within the error bars. Furthermore, we also estimate the difference between the two distributions with the K-S test. It is found that the probability of the K-S test is $0.002$, suggesting that the difference cannot be statistically claimed.

\section{Comparison of the Characteristics of newborn Magnetars in the sGRBs and lGRBs}

\subsection{Braking index}
The temporal evolution of the injected kinetic luminosity of dipole radiations is sensitive to the braking index and characteristic spin down timescale of the magnetar (e.g. Lasky et al. 2017; L\"{u} et al. 2019), which is given by
\begin{eqnarray}
L_{k}(t)\propto
(1+\frac{t}{\tau})^{\frac{4}{1-n}}.
\end{eqnarray}
We derive the braking index $n$ by fitting the XRT lightcurve of the DR component with Eq (1) for employing a Markov chain Monte Carlo (MCMC) algorithm. It would be worth pointing out that the $n$ value derived from our fit has degenerated with $\tau$ to some extent. Taking GRB 130603B as an example, we show the probability contours of the parameters derived from our MCMC fit in Figure \ref{n}. One can find that the contours in the $n-\tau$ plain are prolate, indicating that the two quantities are covariant to some extent. The large uncertainties of $\tau$ and $n$ are due to their covariance.

Our results for the sGRBs are reported in Table 2. Figure \ref{derived parameters}(a) shows the $n$ distribution of the sGRBs in comparison with the lGRBs taken from Zou et al. (2019). The $n^{s}$ distribution ranges from $1.91 \pm 1.1$ to $3.55 \pm 0.84$, and can be fitted with a Gaussian function, i.e. $n^{s}=2.66\pm 0.55$ with an adjusted $R^{2}=0.96$. The $n^{l}$ distribution peaks at around $4$, but has a shoulder extending down to around 2. A tentative $n^{l}$ bimodal distribution with peaks at 2.81 and 4.08 is claimed with a chance probability of $p\sim 10^{-3}$ by estimating the bimodality with the KMM algorithm (Ashman et al. 1994), although the hypothesis of a single Gaussian component in the $n^{l}$ distribution cannot be statistically rejected with a chance probability of $p<10^{-4}$. Twelve lGRBs are assigned to the small-$n^{l}$ group, and 34 lGRBs are allocated to the large-$n^{l}$ group. Our fit to the $n^{l}$ distribution with a two-Gaussian functiongives $n_1^{l}=2.93\pm 0.51$ and $n_2^{l}=4.11\pm 0.39$ ($R_a^{2}=0.94$). The small-$n^{l}$ component is roughly consistent with the $n^{s}$ distribution. In case that the DRs dominate the magnetar spin-down, one has $n=3$ and $L_{k}(t)\propto (1+t/\tau)^{-2}$. If the magnetar spin-down is dominated by the gravitational wave (GW) emission, it is $n=5$ and $L_{k}(t)\propto (1+t/\tau)^{-1}$. One can find that the spin-down of the newborn magnetars in the sGRBs is dominated by the DRs, and the GW emission cannot be ignored for a large fraction (34/46) of the magnetars for making them spin down ($3<n<5$). We should note that the $n$ value may not be strictly equal to 3 in this scenario. The alignment and magnetic dipole moment decline of a newborn magnetar in the early stage may result in the braking index evolution, and its value can be greater than 3 (\c{S}a\c{s}maz Mu\c{s} et al. 2019). The $n$ value may also be slightly greater than 3 when the radiation efficiency decreases with time (Xiao \& Dai 2019). Alternately, Metzger et al. (2018) suggested that the fall-back accretion may lead to $n<3$. We do not consider the temporal evolution of $n$ in this analysis.

\subsection{Initial Spin Period and Magnetic Field Strength}
Initial spin period ($P_0$) and surface polar cap magnetic field strength ($B_{p}$) are critical physical properties of a newborn magnetar. In the scenario that the DR dominates the spindown of a magnetar, its $P_0$  and $B_p$ can be estimated with its DRs. We estimate the $P_0$  and $B_p$ values for those sGRBs and lGRBs in our sample with $n<3$ or $n\sim 3$ within its error bars (e.g. Zhang \& M\'{e}sz\'{a}ros 2001; L\"{u} et al. 2018),
\begin{eqnarray}
B_{\rm p,15} = 2.05(I_{45} R_6^{-3} L_{k,49}^{-1/2} \tau_{3}^{-1})~\rm G,\\
P_{0,-3} = 1.42(I_{45}^{1/2} L_{k,49}^{-1/2} \tau_{3}^{-1/2})~\rm s.
\label{Bp-tau}
\end{eqnarray}
where $Q_n$ is $Q/10^{n}$ in the cgs units, $I$ is the moment of inertia, and $R$ is the radius of the magnetar. We take $I=10^{45}$ g cm$^{2}$ and $R=10^{6}$ cm for our analysis.

Defined the radiation efficiency ($\eta$) of the DR wind as $\eta=L_b/L_k$, $L_k$ then is calculated with $L_k=L_b/\eta$. The radiation efficiency is quite uncertain. Xiao \& Dai (2019) showed $\eta$ as a function of $L_k$ for different bulk saturation Lorentz factors ($\Gamma_{\rm sat}$) of the DR wind. The efficiency dramatically increases with the decrease of the $\Gamma_{\rm sat}$ value. In the case of $\Gamma_{\rm sat}<100$ and even smaller, the efficiency is possibly closer to 0.1. Rowlinson et al. (2014) suggested that the efficiency of the conversion of rotational energy from the magnetar into the observed plateau luminosity is $\leq 20\%$. Du et al. (2016) estimated the radiation efficiency as ~0.3 for GRB 070110. The available observational data cannot place constraints on $\eta$ for each GRBs in our sample. We therefore take three values of $\eta$ for evaluating our results, i.e., $\eta=0.3, 0.1, 0.01$.

The derived $B_p$ and $P_0$ values for the sGRBs sample are presented in Table 2. The distributions of the $B_p$ and $P_0$ for $\eta= 0.3$ are shown in Figure \ref{derived parameters}(b) and \ref{derived parameters}(c). It is found that the $B^{s}_p$ distribution ranges in $8\times10^{14}\sim 4\times10^{16}$ G. It can be fitted with a Guassian function, i.e. $\log B^{s}_p/{\rm G}=16.07\pm 0.34$ with $R_a^2=0.87$. The distribution of $P^{s}_0$ is broad and can be fitted with $\log P^{s}_{0}/{\rm ms}=1.25\pm 0.41$ with an adjusted $R^2$ value as 0.74. For the lGRBs, their $\log B^{l}_p$ normally distributes at $\log B^{l}_p/{\rm G}=14.98\pm 0.51$ with $R_a^2 = 0.78$, and the $P^{l}_0$ narrowly clusters at $\log P^{l}_{0}/{\rm ms}=0.22\pm 0.31$ with $R_a^2 = 0.85$. The mean $B^{s}_p$ of the magnetars is one order magnitude higher than that of $B^{l}_p$, and the mean of $P^{s}_0$ is also one order of magnitude larger than that of the $P^{l}_0$. In the case of $\eta=0.1$, one has $\log (B^{s}_p/{\rm G})=15.80\pm 0.32$ and $\log (B^{l}_p/{\rm G})=14.70\pm 0.35$. In this case, a large fraction of $P^{l}_0$ values breaks the lower limit of 0.98 ms for a neutron star (Lattimer \& Prakash 2004). In the case of $\eta=0.01$, the distribution of $B_p$ of the magnetars in both the sGRBs and lGRBs is similar to that in the case of $\eta=0.3$. All $P^{l}_0$ values are shorter than the breakup spin period limit in this case. Thus, a low radiation efficiency, such as $\eta<0.01$, may challenge the current NS model, although the breakup limit of a NS depends on the equation of state and the mass of the star theoretically (Lattimer \& Prakash 2004) and it is not confirmed observationally.

Figure \ref{p-n} shows $P_0$ as a function of $E_{\rm jet}$ and $n$ in the case of $\eta=0.3$. Both the sGRBs and lGRBs shape a sequence. The sGRBs are in the larger $P_0$, lower $E_{\rm jet}$ end. Our best Spearman correlation analysis for the combined sample of both the sGRBs and lGRBs gives a linear correlation coefficient of $r=-0.91 \pm 0.01$ and $p\sim 5.36\times10^{-10}$. By considering the observational uncertainty of the data, our best linear fit yields $\log P_0/{\rm ms} =(20.60 \pm 1.96)-(0.39\pm 0.04)\log E_{\rm jet}/{\rm ergs}$. In cases of $\eta=0.01$ and 0.1, the correlation is till $ P_0\propto E^{-(0.39\pm 0.04)}_{\rm jet}$. No statistical correlation between $n$ and $P_0$ for the lGRBs and sGRBs is found, as shown in Figure \ref{p-n}(b).

\section{Discussion}

\subsection{Magnetic field strength and Jet formation in Newborn Magnetars}
As shown above, the derived magnetic field strength of the magnetars in the sGRBs is typically $10^{16}$ G, assuming that $R=10^{6}$ cm, $I=10^{45}$ g cm$^{2}$, and a radiation efficiency of the DR wind of 0.3. By performing fully general relativistic magnetohydrodynamic simulations, Giacomazzo \& Perna (2013) showed that a stable magnetar can be formed in a binary NS merger, and its magnetic field is amplified by about two orders of magnitude in their global simulations, even further increased up to magnetar levels in local simulations\footnote{The global simulation means the full general relativistic magnetohydrodynamic simulations of binary neutron star mergers, while the local simulation means the small scale dynamics and other instabilities such as the magnetorotational instability (Giacomazzo \& Perna 2013).}. To properly resolve the instabilities and magnetic field amplification, Giacomazzo et al. (2015) made simulations with subgrid models for binary NS mergers. They showed that the maximum of the magnetic field saturates up to $\sim10^{17}$ G and the mean value saturate is $\sim10^{16}$ G. Such large magnetic fields can lead to the production of a collimated magnetic field aligned with its spin axis, and a relativistic jet should be ejected to power a sGRB (e.g. Meier et al. 2001; Duez et al. 2006; Kiuchi et al. 2012; Siegel et al. 2013, 2014, 2018), as well as a long-lasting GW emission (Dall'Osso et al. 2015), as observed in GRB 170817A/GW 170817 (Abbott et al. 2017a, b).

Although the amplification mechanism of strong magnetic field strength is debated, time-dependent axisymmetric magnetohydrodynamic simulations by Bucciantini et al. (2009) showed that a magnetar in collapses of a massive star may be formed. We find that the $B_p$ values of the magnetars in the lGRBs are averagely smaller than that in the sGRB by about one order of magnitude. Superluminous supernovae (SLSNe) are very similar to hypernovae associated with lGRBs (van den Heuvel \& Portegies Zwart 2013; Lunnan et al. 2015; Nicholl et al. 2015, 2016; Japelj et al. 2016; Jerkstrand et al. 2017; Yu et al. 2017; Prajs et al. 2017; Inserra et al. 2018). In particular, Lunnan et al. (2015) showed that the host galaxies of GRBs and SLSNe Type I share many common properties, e.g. high star-formation rate and low metallicity, although some differences still exist (Angus et al. 2016). They may also be driven by magnetars (Maeda et al. 2007; Kasen \& Bildsten 2010; Woosley 2010; Wang et al. 2015; Yu et al. 2017; Liu et al. 2017). Liu et al. (2017) collected 19 hydrogen-deficient SLSNe (Type I) and fitted their lightcurves, temperature evolution, and velocity evolution based on the magnetar-powered model. Yu et al. (2017) fitted the bolometric lightcurves of 31 SLSNe with the magnetar engine model to derive the magnetar parameters and compared these parameters with the magnetars in lGRBs (see also Nicholl et al. 2017). We add $B_p$ and $P_0$ distributions of these SLSNe, which are taken from Yu et al. (2017), into Figure \ref{derived parameters} (b) and \ref{derived parameters} (c). It is found that the typical $B_p$ value of the magnetars in SLSNe is $10^{14}$ G, being one and two orders of magnitude lower than the typical values of the magnetars in the lGRBs and sGRBs, respectively. This may be due to the selection effect since these magnetars are selected with their strong jets, saying, bright GRBs. However, non-detection of association between SLSNe and GRBs hints that the magnetic field strength is critical for jet formation in death of massive stars. Time-dependent axisymmetric magnetohydrodynamic simulations by Bucciantini et al. (2009) showed that a newborn magnetar in collapses of a massive star may drive a jet within its strong toroidal magnetic field, and they also showed that most of the spin down power of the magnetar escapes via the relativistic jet. As shown in Figure \ref{p-n} (a), the jet energy powered by a magnetar is likely correlated with $P_0$. The $P_0$ distribution of the magnetars in SLSNe is not statistically different from that of the lGRBs. One can expect that the jet energy of the SLSNe, if they are successfully launched, should be similar to the lGRB jets. The non-detection of jet emission accompanied by these SLSNe would be due to failure of the jet launch since a weak jet may be choked by extended material and stalled sufficiently far below the photosphere (Senno et al. 2016), although one still cannot rule out the possibility that the non-detection of the GRB jet accompanied by these SLSNe is due to the misalignment effect of the jet to the light of sight.

\subsection{Spin Down Mechanisms of Newborn Magnetars}
Our analysis shows that $n\sim 3$ {\em vs.} $n\sim 4$ and $P_0\sim 20$ ms {\em vs.} $P_0\sim 2$ ms in comparison of the newborn magnetars in the sGRBs {\em vs.} lGRBs. The difference in the braking indices implies that the spindown tends to be dominated by the magnetic DRs for the magnetars in the sGRBs, while it is co-operated by both the magnetic DRs and the GW emission for the magnetars in the lGRB. We note that these magnetars should experience dramatically different evolution in their early stage. For example, the alignment and magnetic dipole moment decline and the fall-back accretion of a newborn magnetar should result in the evolutions of the braking index and spin period in the early stage (e.g. M\'{e}sz\'{a}ros 2001; Fan et al. 2013; Giacomazzo \& Perna 2013; Lasky \& Glampedakis 2016; L\"{u} et al. 2018; Metzger et al. 2018; \c{S}a\c{s}maz Mu\c{s} et al. 2019).

The difference between the magnetars born in merger of compact stars and collapse of massive stars should stem from the properties of their progenitors and the dynamical behaviors of the magnetar formation. The dynamical behavior of the merger of binary NSs depends on the equation of state (EoS) and the total mass of the binary (Hotokezaka et al. 2013). The outcomes of the mergers may be a BH (Rosswog et al. 2003; Rezzolla et al. 2011), a differential-rotation-supported hypermassive NS (HMNS) that lasts for $\sim100$ ms before collapsing to a BH (Rosswog et al. 2003; Shibata \& Taniguchi 2006; Sekiguchi et al. 2011), a rigid-rotation-supported supra-massive NS (SMNS) that can survive for a much longer time (e.g. tens of seconds to $>10^{4}$ s) before collapsing (Shibata \& Shapiro 2002; Lasky et al. 2014; L\"{u} et al. 2015; Breu \& Rezzolla 2016), and a stable magnetar (Giacomazzo \& Perna 2013). The selection of the GRBs in our sample excludes events that are relevant to the first and second scenarios. 8 events in our sGRBs have an internal X-ray plateau, which is regarded as an indicator of collapses of a magnetar into a black hole. Their $t^{s}_b$ values range from 174 seconds to 594 seconds (see Table 1). These events are consistent with the third scenario, i.e. a rigid-rotation-supported supra-massive NS (SMNS) as the outcome of the mergers. The magnetars in the other sGRBs may be stable.

As discussed above, the typical magnetic field strength of the magnetars in the sGRBs is stronger than that in the lGRBs. Thus, the magnetic DRs may dominate the spin-down of these magnetars. Normal stable pulsars, due to their long-term spindown, are believed to be DR dominated (Antonopoulou et al. 2015; Archibald et al. 2016; Clark et al. 2016). We show that the braking index distribution with a sample of stable pulsars from L\"{u} et al. (2019)\footnote{Two pulsars of the same have a braking index much smaller than 3, i.e., $n$ of $1.4\pm0.2$ for Vela (PRS B0833-45) and  $n=0.9\pm0.2$ for PRS J1734-3333. They may have the rotational properties of a magnetar (Lyne et al. 1996; Espinoza et al. 2011).} in Figure \ref{derived parameters}(a). It is consistent with the $n^{s}$ distribution. This suggests that the spin down mechanism of the magnetars in the sGRBs is similar to stable pulsars.

\section{Conclusions}
Using the data observed with the {\em Swift} BAT and XRT, we have presented a comparative analysis between the sGRBs and lGRBs whose early X-ray afterglow lightcurves are composed of a shallow decay segment, and explored the characteristics of magnetars born in death of massive stars and merger of compact objects assuming that the shallow decay segment is attributed to the DRs of the magnetars. We show that the isotropic gamma-ray/X-ray energies of the jets and DR winds of the sGRBs are averagely lower than that of lGRBs by about two orders of magnitudes. Their spectra are systematically harder than the lGRBs, but the spectra of the DR winds in the sGRBs are not statistically different from that of the lGRBs; 8 events among 27 sGRBs in our sample have an internal X-ray plateau, which lasts from 174 s to 594 s, being consistent with a rigid-rotation-supported supra-massive magnetar as the outcome of the compact NS mergers.

Deriving the magnetar braking index by fitting the X-ray lightcurves without considering the temporal evolution, we find that the $n^{l}$ distribution illustrates as a bimodal structure peaking at 2.93 and 4.11, with a number ratio of these two components is 12:34. The $n^{s}$ distribution can be fitted with a Gaussian function of $n^{s}=2.66\pm 0.55$, which is consistent with the low-$n$ component of the $n^{l}$ distribution and the $n$ distribution of old stable pulsars. By selecting a sub-sample of the magnetars whose spin-down is DR dominated with a criterion of $n\sim 3$ and adopting a universal radiation efficiency of $\eta=0.3$, we find that $\log B^{s}_p/{\rm G}=16.07\pm 0.34$, which is about one order of magnitude higher than that of the lGRBs, i.e. $\log B^{l}_p/{\rm G}=14.98\pm 0.51$. The $\log B_p$ of the magnetars in the SLSNe is about two orders of magnitude lower than $B^{s}_p$. The typical $\log P_0/{\rm ms}$ values are $1.25\pm 0.41$, $0.22\pm 0.31$, and $0.41\pm 0.20$ for the magnetars in the sGRBs, lGRBs, and SLSNe, respectively. Both the magnetars in the sGRBs and lGRBs follow the same $\log P_0-\log E_{\rm jet}$ relation, i.e. $\log P_0/{\rm ms} =(20.60\pm 1.96)-(0.39\pm 0.04)\log E_{\rm jet}/{\rm ergs}$.

Based on these results, we argue that a magnetar born in merger of compact stars tends to have a stronger $B_p$ and a longer $P_0$ by about one to two orders of magnitude than that born in collapse of massive stars for both the sGRBs and SLSNe, and its spin-down is dominated by the magnetic DRs being due to its strong magnetic field strength, whereas the early spin-down of a magnetar born in a massive star collapses is co-operated by both the DRs and GW emission. A magnetar with a faster rotation speed should power a more energetic jet, being independent of its formation approach.

\section*{Acknowledgements}

We very appreciate helpful comments and suggestions from the referee. We acknowledge the use of the public data from the {\em Swift} data archive and the UK {\em Swift} Science Data Center. This work is supported by the National Natural Science Foundation of China (Grant Nos.12133003, U1731239, and 11922301), Guangxi Science Foundation (Grant No.2017AD22006), and the Program of Bagui Young Scholars Program (LHJ), and Innovation Project of Guangxi Graduate Education (Grant No.YCBZ2020025).

%**********************************************************************

\section*{Data availability}

The data underlying this article are available in the article.

%*******************************************************************************************
%*******************************************************************************************
\clearpage

\begin{landscape}
\begin{table}
%\centering
\caption{Data of the sGRBs in our sample}
\begin{minipage}{155mm}
\begin{tabular}{lllllllllllll}
\hline
GRB & z & $\alpha_{1}^{a}$ & $\alpha_{2}^{a}$ & $t_{\rm b,2}^{a}$ & $\Gamma_{\rm jet}^{b}$ & $\Gamma_{\rm wind}^{b}$ & $S_{\rm jet,-7}^{c}$ & $S_{\rm w,-7}^{c}$ & $L_{\rm b,46}^{d}$ & $E_{\rm jet,50}^{e}$ & $E_{\rm w,49}^{e}$  \\
\hline
051221A	&	0.5465	&	0.31 	$\pm$	0.11 	&	1.34 	$\pm$	0.07 	&	285.31 	$\pm$	69.22 	&	1.39 	$\pm$	0.06 	&	1.83 	$\pm$	0.21 	&	11.79 	$\pm$	1.30 	&	0.19 	$\pm$	0.02 	&	0.18 	$\pm$	0.04 	&	9.26 	$\pm$	1.02 	&	1.49 	$\pm$	0.28 	&	\\
060313	&		&	0.60 	$\pm$	0.08 	&	1.65 	$\pm$	0.07 	&	59.54 	$\pm$	12.25 	&	0.70 	$\pm$	0.07 	&	1.78 	$\pm$	0.20 	&	11.84 	$\pm$	0.50 	&	0.21 	$\pm$	0.02 	&				&				&				&	\\
060614	&	0.1254	&	0.05 	$\pm$	0.05 	&	1.82 	$\pm$	0.04 	&	457.27 	$\pm$	26.22 	&	2.13 	$\pm$	0.04 	&	1.75 	$\pm$	0.10 	&	217.00 	$\pm$	4.00 	&	0.83 	$\pm$	0.04 	&	$\sim$0.03 	 	&	7.44 	$\pm$	0.12 	&	0.28 	$\pm$	0.02 	&	\\
060801	&	1.31	&	-0.20 	$\pm$	1.16 	&	1.16 	$\pm$	0.27 	&	1.50 	$\pm$	0.74 	&	0.47 	$\pm$	0.23 	&	1.64 	$\pm$	0.37 	&	0.81 	$\pm$	0.10 	&	0.13 	$\pm$	0.03 	&	216.62 	$\pm$	44.19 	&	3.69 	$\pm$	0.46 	&	5.82 	$\pm$	3.53 	&	\\
061201	&	0.111	&	0.62 	$\pm$	0.12 	&	2.12 	$\pm$	0.13 	&	28.82 	$\pm$	6.45 	&	0.81 	$\pm$	0.14 	&	1.30 	$\pm$	0.18 	&	3.30 	$\pm$	0.30 	&	0.72 	$\pm$	0.09 	&	0.16 	$\pm$	0.06 	&	0.10 	$\pm$	0.01 	&	0.21 	$\pm$	0.03 	&	\\
070714A	&		&	0.54 	$\pm$	0.15 	&	1.10 	$\pm$	0.23 	&	51.46 	$\pm$	86.35 	&	2.61 	$\pm$	0.23 	&	2.04 	$\pm$	0.35 	&	1.50 	$\pm$	0.20 	&	0.22 	$\pm$	0.02 	&				&				&				&	\\
070724A	&	0.457	&	0.20 	$\pm$	0.00 	&	2.53 	$\pm$	0.00 	&	412.23 	$\pm$	42.78 	&	1.96 	$\pm$	0.36 	&	1.70 	$\pm$	0.78 	&	0.79 	$\pm$	0.07 	&	0.03 	$\pm$	0.02 	&	0.02 	$\pm$	0.05 	&	0.43 	$\pm$	0.04 	&	0.18 	$\pm$	0.16 	&	\\
070809	&	0.219	&	0.04 	$\pm$	0.18 	&	1.17 	$\pm$	0.28 	&	71.06 	$\pm$	37.35 	&	1.70 	$\pm$	0.23 	&	1.48 	$\pm$	0.32 	&	1.03 	$\pm$	0.10 	&	0.09 	$\pm$	0.01 	&	0.04 	$\pm$	0.01 	&	0.12 	$\pm$	0.01 	&	0.10 	$\pm$	0.02 	&	\\
080426	&		&	-0.18 	$\pm$	0.82 	&	1.30 	$\pm$	0.05 	&	4.38 	$\pm$	1.15 	&	1.97 	$\pm$	0.13 	&	1.39 	$\pm$	0.49 	&	3.70 	$\pm$	0.30 	&	0.08 	$\pm$	0.01 	&				&				&				&	\\
080702A	&		&	0.51 	$\pm$	0.15 	&	3.60 	$\pm$	0.47 	&	5.94 	$\pm$	1.06 	&	1.24 	$\pm$	0.46 	&	1.40 	$\pm$	0.79 	&	0.36 	$\pm$	0.10 	&	0.06 	$\pm$	0.02 	&				&				&				&	\\
080919	&		&	0.22 	$\pm$	0.61 	&	9.39 	$\pm$	17.82 	&	4.57 	$\pm$	3.58 	&	1.10 	$\pm$	0.25 	&	2.72 	$\pm$	0.82 	&	0.72 	$\pm$	0.11 	&	0.73 	$\pm$	0.06 	&				&				&				&	\\
090426	&	2.6	&	-0.40 	$\pm$	0.88 	&	1.05 	$\pm$	0.11 	&	2.23 	$\pm$	0.86 	&	1.93 	$\pm$	0.23 	&	1.81 	$\pm$	0.34 	&	1.80 	$\pm$	0.30 	&	0.05 	$\pm$	0.01 	&	252.27 	$\pm$	68.14 	&	28.43 	$\pm$	4.74 	&	8.18 	$\pm$	6.62 	&	\\
090510	&	0.903	&	0.70 	$\pm$	0.04 	&	2.26 	$\pm$	0.11 	&	16.53 	$\pm$	2.47 	&	0.95 	$\pm$	0.16 	&	1.60 	$\pm$	0.12 	&	3.40 	$\pm$	0.40 	&	1.48 	$\pm$	0.13 	&	38.09 	$\pm$	6.76 	&	7.45 	$\pm$	0.88 	&	32.53 	$\pm$	5.48 	&	\\
090515	&		&	-0.26 	$\pm$	0.38 	&	2.41 	$\pm$	0.44 	&	1.36 	$\pm$	0.17 	&	1.41 	$\pm$	0.24 	&	1.70 	$\pm$	0.16 	&	0.21 	$\pm$	0.04 	&	0.65 	$\pm$	0.07 	&				&				&				&	\\
100117A	&	0.92	&	0.67 	$\pm$	0.16 	&	4.53 	$\pm$	0.17 	&	2.27 	$\pm$	0.15 	&	0.88 	$\pm$	0.22 	&	1.45 	$\pm$	0.12 	&	0.93 	$\pm$	0.13 	&	0.90 	$\pm$	0.07 	&	224.20 	$\pm$	36.00 	&	2.12 	$\pm$	0.30 	&	20.53 	$\pm$	3.14 	&	\\
100625A	&	0.425	&	0.35 	$\pm$	0.59 	&	3.16 	$\pm$	1.02 	&	2.36 	$\pm$	0.72 	&	0.31 	$\pm$	0.91 	&	2.03 	$\pm$	0.32 	&	8.30 	$\pm$	1.50 	&	0.06 	$\pm$	0.02 	&	1.82 	$\pm$	0.99 	&	3.87 	$\pm$	0.70 	&	0.28 	$\pm$	0.15 	&	\\
100702A	&		&	0.69 	$\pm$	0.12 	&	5.31 	$\pm$	0.31 	&	2.13 	$\pm$	0.07 	&	1.54 	$\pm$	0.15 	&	1.90 	$\pm$	0.13 	&	1.20 	$\pm$	0.10 	&	1.57 	$\pm$	0.58 	&				&				&				&	\\
100724A	&	1.288	&	0.71 	$\pm$	0.17 	&	1.57 	$\pm$	0.19 	&	43.81 	$\pm$	28.69 	&	1.93 	$\pm$	0.21 	&	1.94 	$\pm$	0.27 	&	1.60 	$\pm$	0.20 	&	0.17 	$\pm$	0.03 	&	5.28 	$\pm$	4.20 	&	7.05 	$\pm$	0.88 	&	7.40 	$\pm$	2.78 	&	\\
101219A	&	0.718	&	0.23 	$\pm$	0.69 	&	6.24 	$\pm$	1.91 	&	1.74 	$\pm$	0.23 	&	0.63 	$\pm$	0.10 	&	1.34 	$\pm$	0.32 	&	4.60 	$\pm$	0.30 	&	0.34 	$\pm$	0.09 	&	87.78 	$\pm$	31.24 	&	6.34 	$\pm$	0.41 	&	4.66 	$\pm$	2.08 	&	\\
120305A	&		&	-0.03 	$\pm$	0.27 	&	7.07 	$\pm$	0.40 	&	1.89 	$\pm$	0.06 	&	1.00 	$\pm$	0.09 	&	1.90 	$\pm$	0.25 	&	2.13 	$\pm$	0.10 	&	0.88 	$\pm$	0.09 	&				&				&				&	\\
120521A	&		&	0.31 	$\pm$	0.25 	&	10.65 	$\pm$	9.28 	&	2.71 	$\pm$	0.46 	&	0.98 	$\pm$	0.22 	&	1.56 	$\pm$	0.24 	&	0.78 	$\pm$	0.11 	&	0.15 	$\pm$	0.02 	&				&				&				&	\\
130603B	&	0.356	&	0.42 	$\pm$	0.07 	&	1.68 	$\pm$	0.07 	&	31.31 	$\pm$	5.97 	&	0.82 	$\pm$	0.07 	&	1.77 	$\pm$	0.20 	&	6.30 	$\pm$	0.30 	&	0.57 	$\pm$	0.06 	&	1.62 	$\pm$	0.42 	&	2.03 	$\pm$	0.10 	&	1.82 	$\pm$	0.24 	&	\\
130912A	&		&	0.52 	$\pm$	0.36 	&	1.55 	$\pm$	0.09 	&	4.34 	$\pm$	1.95 	&	1.20 	$\pm$	0.20 	&	1.31 	$\pm$	0.36 	&	1.70 	$\pm$	0.20 	&	0.43 	$\pm$	0.08 	&				&				&				&	\\
140903A	&	0.351	&	0.07 	$\pm$	0.07 	&	1.22 	$\pm$	0.08 	&	85.20 	$\pm$	15.21 	&	1.99 	$\pm$	0.12 	&	1.60 	$\pm$	0.21 	&	1.54 	$\pm$	0.10 	&	0.23 	$\pm$	0.03 	&	0.52 	$\pm$	0.07 	&	0.48 	$\pm$	0.03 	&	0.73 	$\pm$	0.13 	&	\\
150120B	&		&	0.15 	$\pm$	0.31 	&	1.19 	$\pm$	0.06 	&	13.48 	$\pm$	4.52 	&	2.20 	$\pm$	0.26 	&	1.87 	$\pm$	0.26 	&	5.40 	$\pm$	0.80 	&	0.24 	$\pm$	0.03 	&				&				&				&	\\
160525B	&		&	-0.54 	$\pm$	1.74 	&	1.98 	$\pm$	0.29 	&	2.94 	$\pm$	0.84 	&	1.88 	$\pm$	0.36 	&	2.97 	$\pm$	0.89 	&	0.34 	$\pm$	0.08 	&	0.06 	$\pm$	0.01 	&				&				&				&	\\
160927A	&		&	-0.23 	$\pm$	1.31 	&	0.99 	$\pm$	0.05 	&	1.36 	$\pm$	0.46 	&	0.36 	$\pm$	0.26 	&	1.73 	$\pm$	0.68 	&	1.40 	$\pm$	0.20 	&	0.01 	$\pm$	0.01 	&				&				&				&	\\
190627A	&	1.942	&	0.09 	$\pm$	0.11 	&	1.51 	$\pm$	0.09 	&	213.57 	$\pm$	38.83 	&	1.64 	$\pm$	0.05 	&	1.99 	$\pm$	0.16 	&	0.99 	$\pm$	0.22 	&	0.38 	$\pm$	0.03 	&	37.13 	$\pm$	7.42 	&	9.37 	$\pm$	2.08 	&	35.50 	$\pm$	7.66 	&	\\
\hline
\end{tabular}
Note: \\
${a}$---$\alpha_{1}$ and $\alpha_{2}$ are the decay slope before and after the break time. $t_{b}$ is the break time of lightcurves from our fitting, in units of $10^{2}$ s. \\
${b}$---The spectral photon index of the magnetar jet and wind.\\
${c}$---The fluences of the jet and DR wind, in units of $10^{-7}$ erg cm$^{-2}$.\\
${d}$---The plateau luminosity of our fits, in units of $10^{46}$ erg s$^{-1}$.\\
${e}$---The isotropic energy releases of the jet and DR wind, in units of $10^{50}$ erg and$10^{49}$ erg, respectively.
\end{minipage}
\end{table}
\end{landscape}
%\end{center}

%*******************************************************************************************
\clearpage
\begin{table}
\caption{The derived parameters of the sGRBs in our samples}
\begin{tabular}{llllllll}
\hline
GRB & $P_{0}^{a}$(ms)& $B_{p,15}^{a}$(G)& $P_{0}^{b}$(ms)& $B_{p,15}^{b}$(G)& $P_{0}^{c}$(ms)& $B_{p,15}^{c}$(G)& $n$ \\
\hline
051221A	&	13.54 	$\pm$	3.12 	&	4.55 	$\pm$	0.52 	&	7.82 	$\pm$	1.80 	&	2.63 	$\pm$	0.30 	&	2.47 	$\pm$	0.57 	&	0.83 	$\pm$	0.09 	&	2.92 	$\pm$	0.27\\
060614	&	23.57 	$\pm$	1.47 	&	5.34 	$\pm$	0.08 	&	13.61 	$\pm$	0.85 	&	3.08 	$\pm$	0.05 	&	4.30 	$\pm$	0.27 	&	0.97 	$\pm$	0.01 	&	2.07 	$\pm$	0.10\\
061201	&	38.38 	$\pm$	11.24 	&	34.40 	$\pm$	4.77 	&	22.16 	$\pm$	6.49 	&	19.86 	$\pm$	2.75 	&	7.01 	$\pm$	2.05 	&	6.28 	$\pm$	0.87 	&	2.74 	$\pm$	0.12\\
070724A	&	35.07 	$\pm$	19.12 	&	9.52 	$\pm$	1.67 	&	20.25 	$\pm$	11.04 	&	5.49 	$\pm$	0.97 	&	6.40 	$\pm$	3.49 	&	1.74 	$\pm$	0.31 	&	1.91 	$\pm$	1.11\\
070809	&	54.00 	$\pm$	22.94 	&	32.29 	$\pm$	9.94 	&	31.18 	$\pm$	13.24 	&	18.64 	$\pm$	5.74 	&	9.86 	$\pm$	4.19 	&	5.90 	$\pm$	1.82 	&	2.65 	$\pm$	0.98\\
090426	&	6.23 	$\pm$	2.01 	&	36.13 	$\pm$	7.25 	&	3.59 	$\pm$	1.16 	&	20.86 	$\pm$	4.18 	&	1.14 	$\pm$	0.37 	&	6.60 	$\pm$	1.32 	&	3.10 	$\pm$	0.55\\
090510	&	4.28 	$\pm$	0.70 	&	6.62 	$\pm$	0.42 	&	2.47 	$\pm$	0.40 	&	3.82 	$\pm$	0.24 	&	0.78 	$\pm$	0.13 	&	1.21 	$\pm$	0.08 	&	2.35 	$\pm$	0.08\\
130603B	&	12.72 	$\pm$	2.81 	&	12.09 	$\pm$	1.17 	&	7.35 	$\pm$	1.62 	&	6.98 	$\pm$	0.67 	&	2.32 	$\pm$	0.51 	&	2.21 	$\pm$	0.21 	&	2.45 	$\pm$	0.24\\
140903A	&	13.59 	$\pm$	2.06 	&	7.81 	$\pm$	0.50 	&	7.85 	$\pm$	1.19 	&	4.51 	$\pm$	0.29 	&	2.48 	$\pm$	0.38 	&	1.43 	$\pm$	0.09 	&	2.69 	$\pm$	0.30 \\
190627A	&	1.50 	$\pm$	0.29 	&	0.80 	$\pm$	0.07 	&	0.86 	$\pm$	0.16 	&	0.46 	$\pm$	0.04 	&	0.27 	$\pm$	0.05 	&	0.15 	$\pm$	0.01 	&	2.24 	$\pm$	0.30\\
\hline
\end{tabular}
\\Note: ${a}$, ${b}$, ${c}$--Three cases of X-ray radiation efficiency of the dipole radiation wind as $\eta=0.3, 0.1, 0.01$, respectively.
\end{table}
\newpage

\begin{figure*}
\includegraphics[angle=0,scale=0.45]{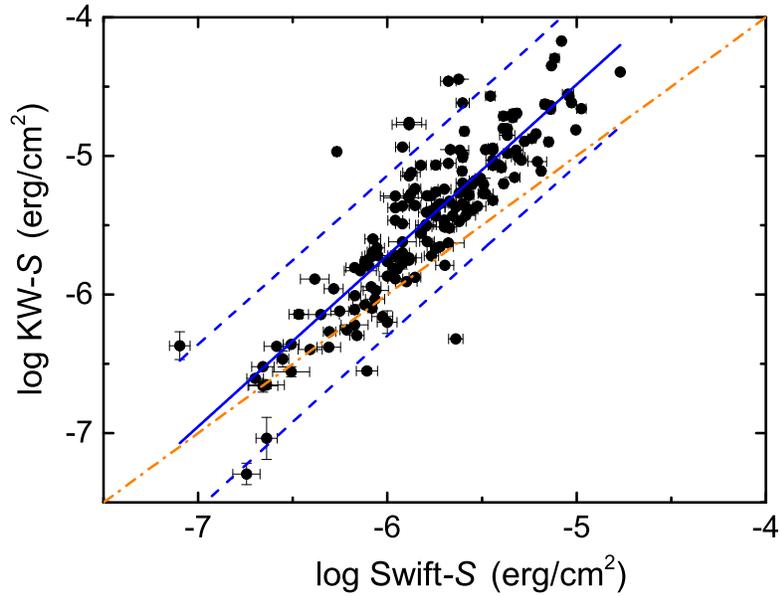}
\center\caption{Comparison of the prompt gamma-ray fluences observed with the {\em Konus}/Wind (20keV -2 MeV) and {\em Swift}/BAT (15-150 keV) for a sample of 168 GRBs simultaneously observed with the two detectors (Tsvetkova et al. 2021). The orange dash-dotted line is the equality line ($\rm y= \rm x$). The red solid and blue dashed lines are the least-squares linear fit and the 95\% confidence level, respectively.}
\label{KW-BAT}
\end{figure*}

\begin{figure*}
\includegraphics[angle=0,scale=0.28]{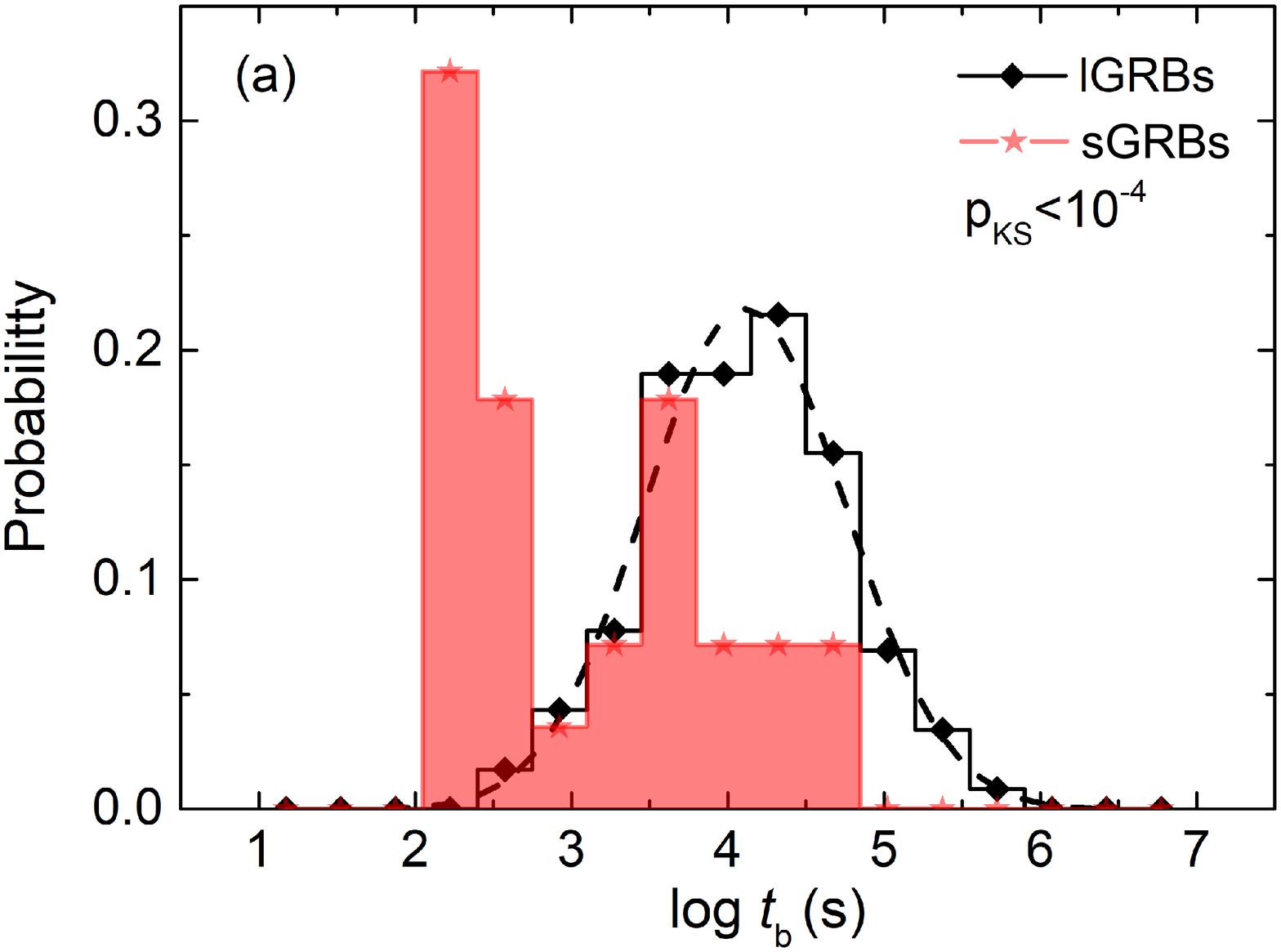}
\includegraphics[angle=0,scale=0.28]{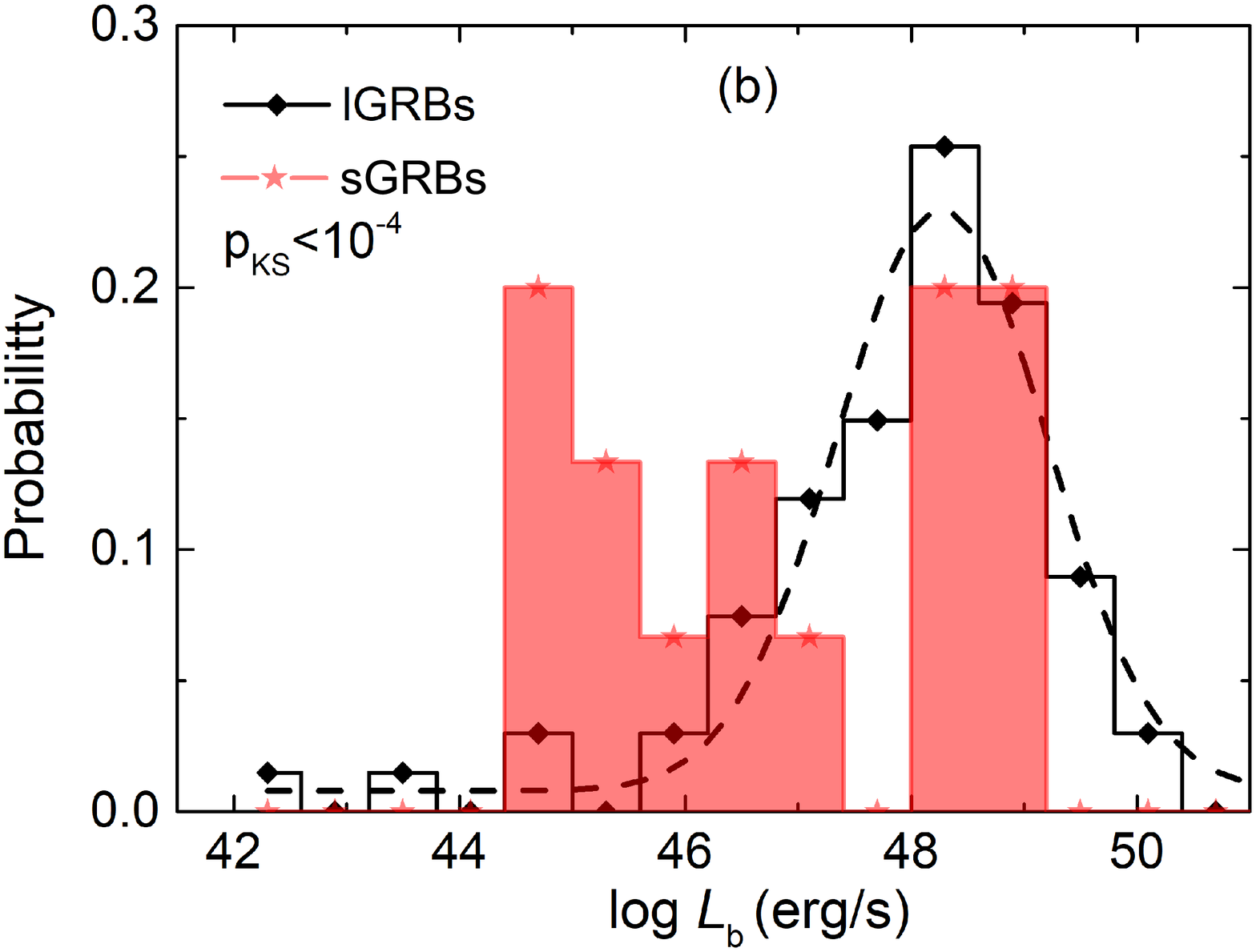}
\includegraphics[angle=0,scale=0.35]{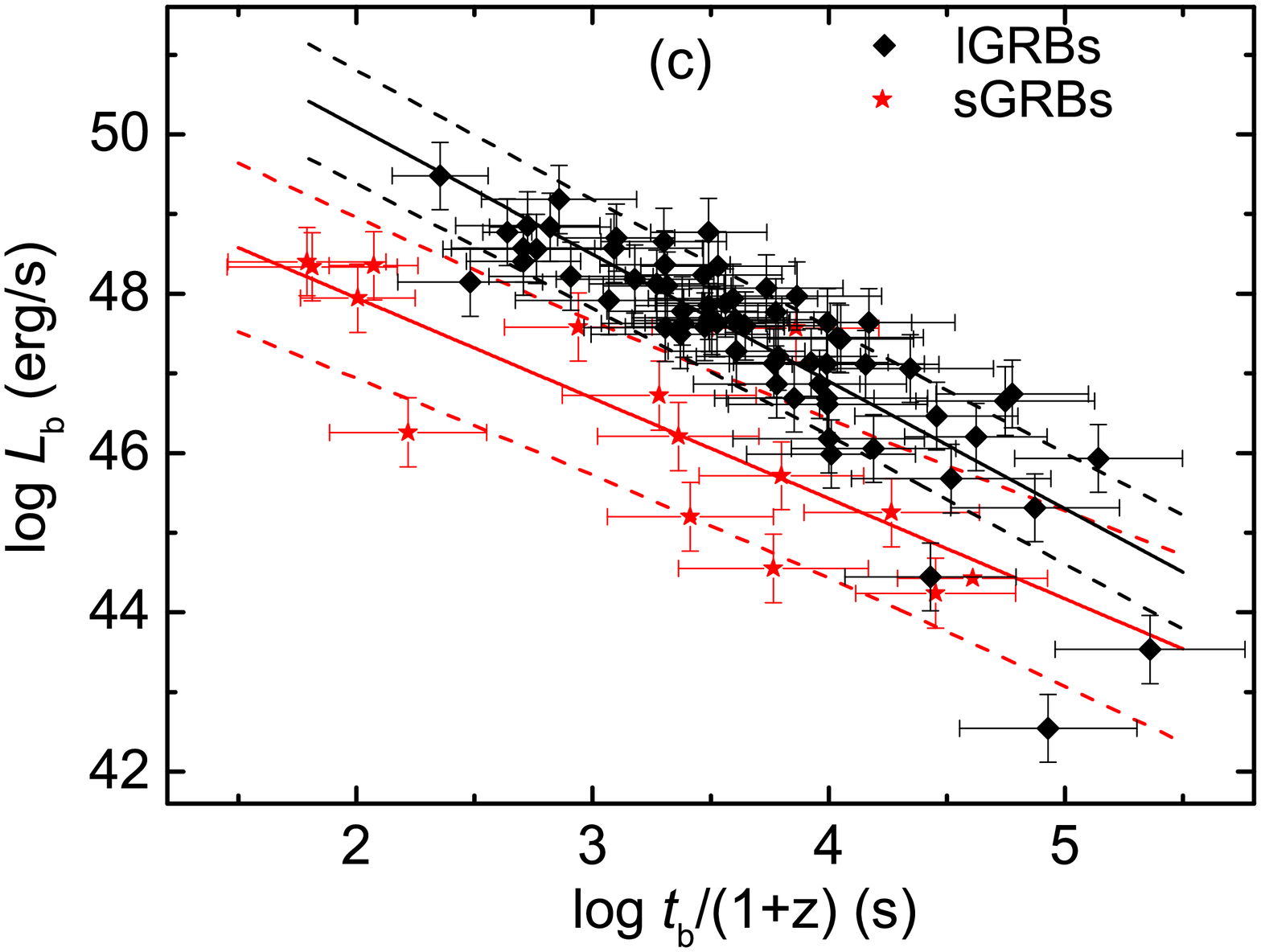}
\center\caption{Panels a and b---Distributions of $t_{b}$ and $L_{b}$ for the sGRBs in our sample in comparison with a sample of lGRBs taken from Zou et al. (2019). The dashed lines are the best Gaussian fits to the corresponding distributions. The probability of the K-S test for examining the difference of the distribution is also marked. Panel c)---$L_b$ as a function of $t_b/(1+z)$ for the sGRBs and lGRBs. The solid and dashed lines are the least-squares linear fit and its 95\% confidence level, respectively.}
\label{tb-Lb}
\end{figure*}

\begin{figure*}
\includegraphics[angle=0,scale=0.28]{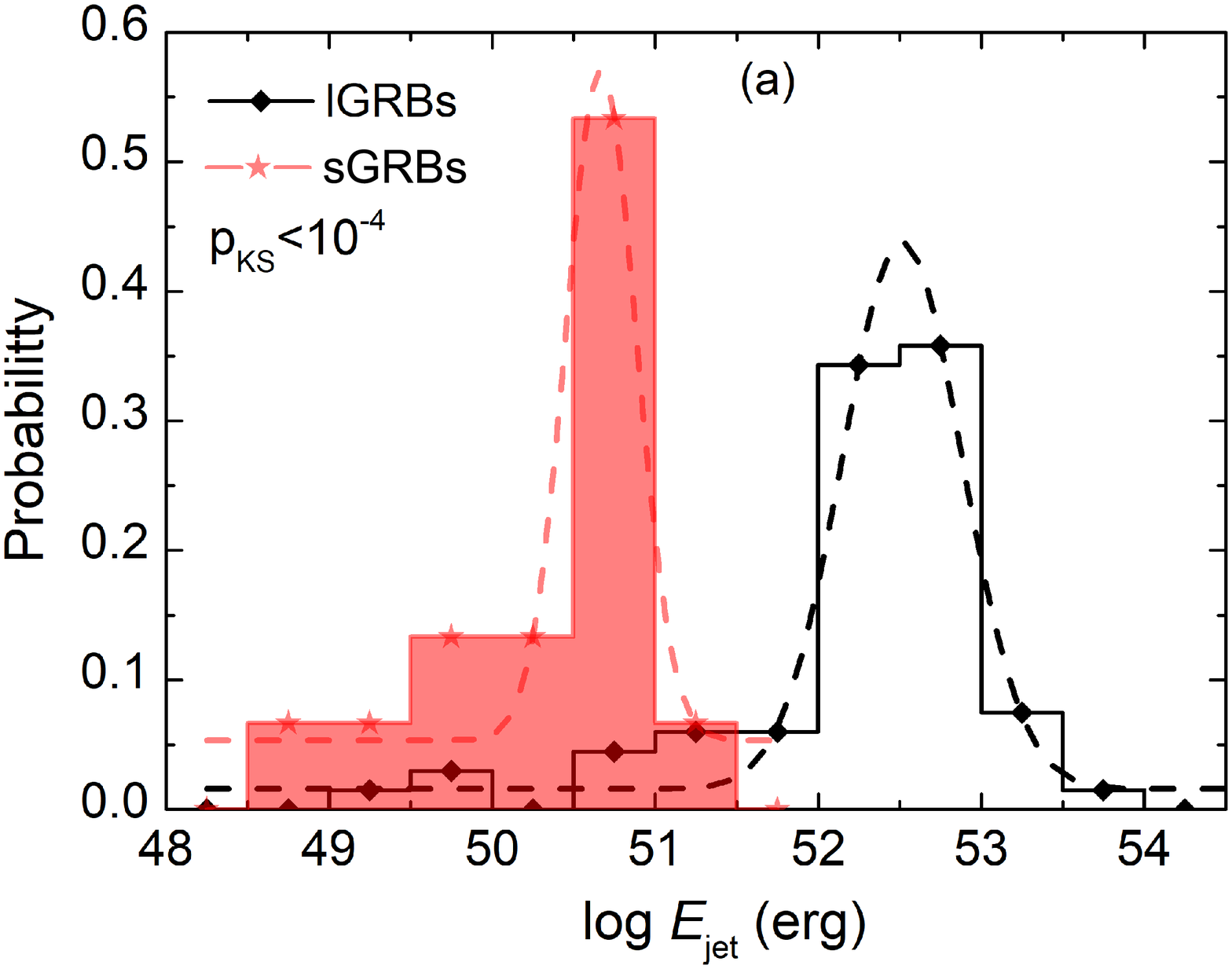}
\includegraphics[angle=0,scale=0.28]{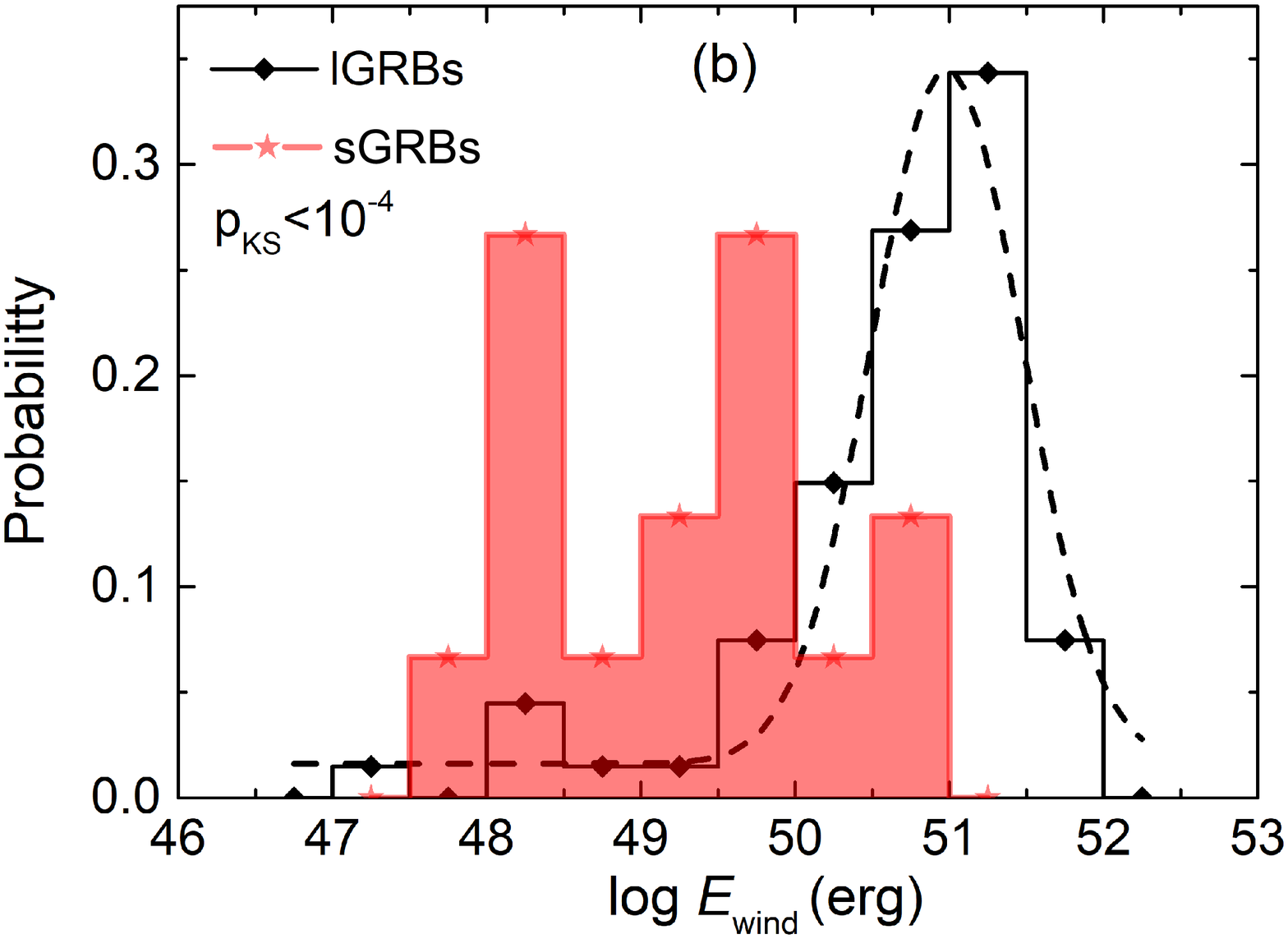}
\center\caption{Panels a and b---Distributions of $E_{\rm jet}$ and $E_{\rm wind}$ for the sGRBs in our sample in comparison with a sample of lGRBs taken from Zou et al. (2019). The dashed lines are the best Gaussian fits to the corresponding distributions. The probability of the K-S test for examining the difference of the distribution is also marked.}
\label{E-E}
\end{figure*}

\begin{figure*}
\includegraphics[angle=0,scale=0.25]{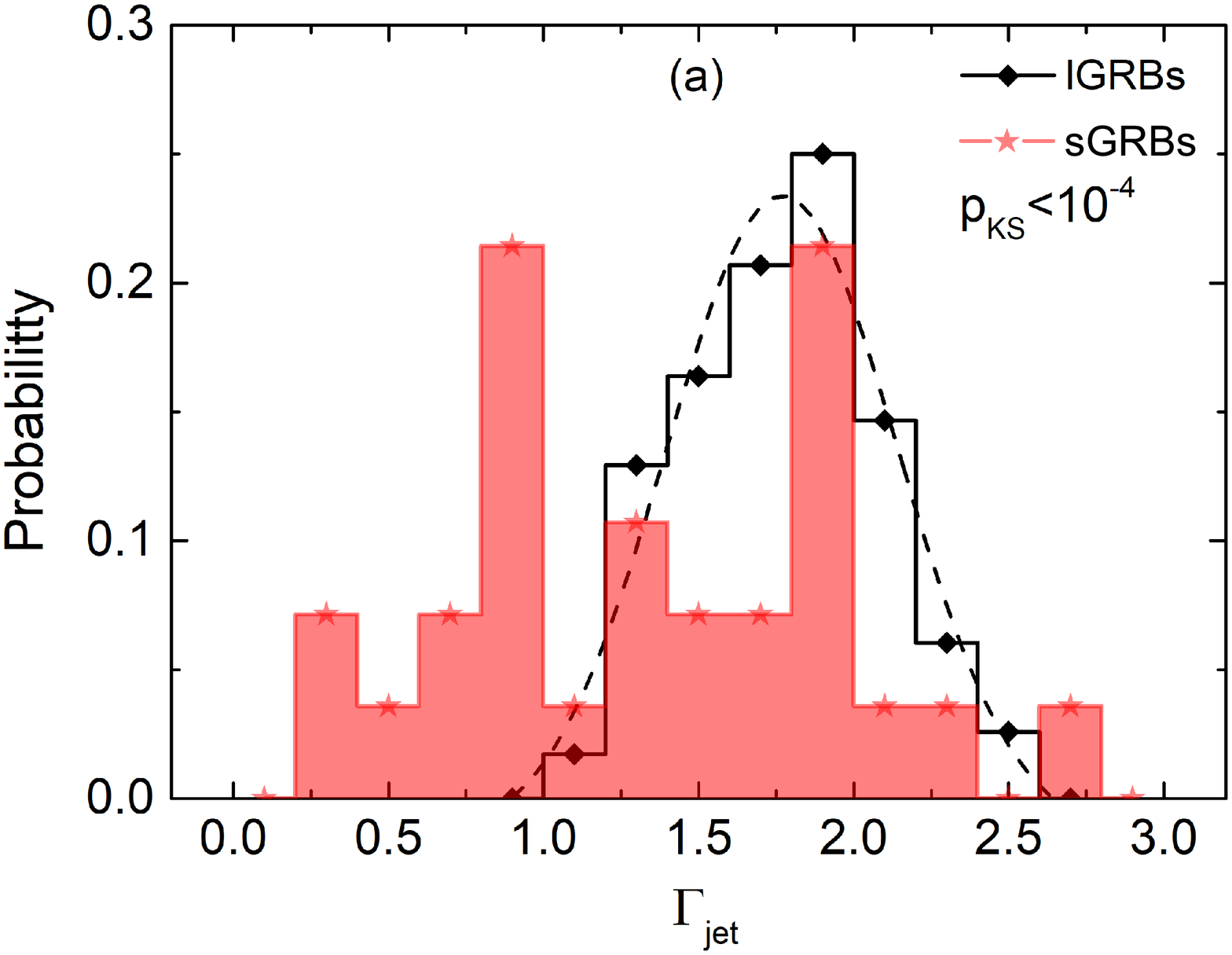}
\includegraphics[angle=0,scale=0.25]{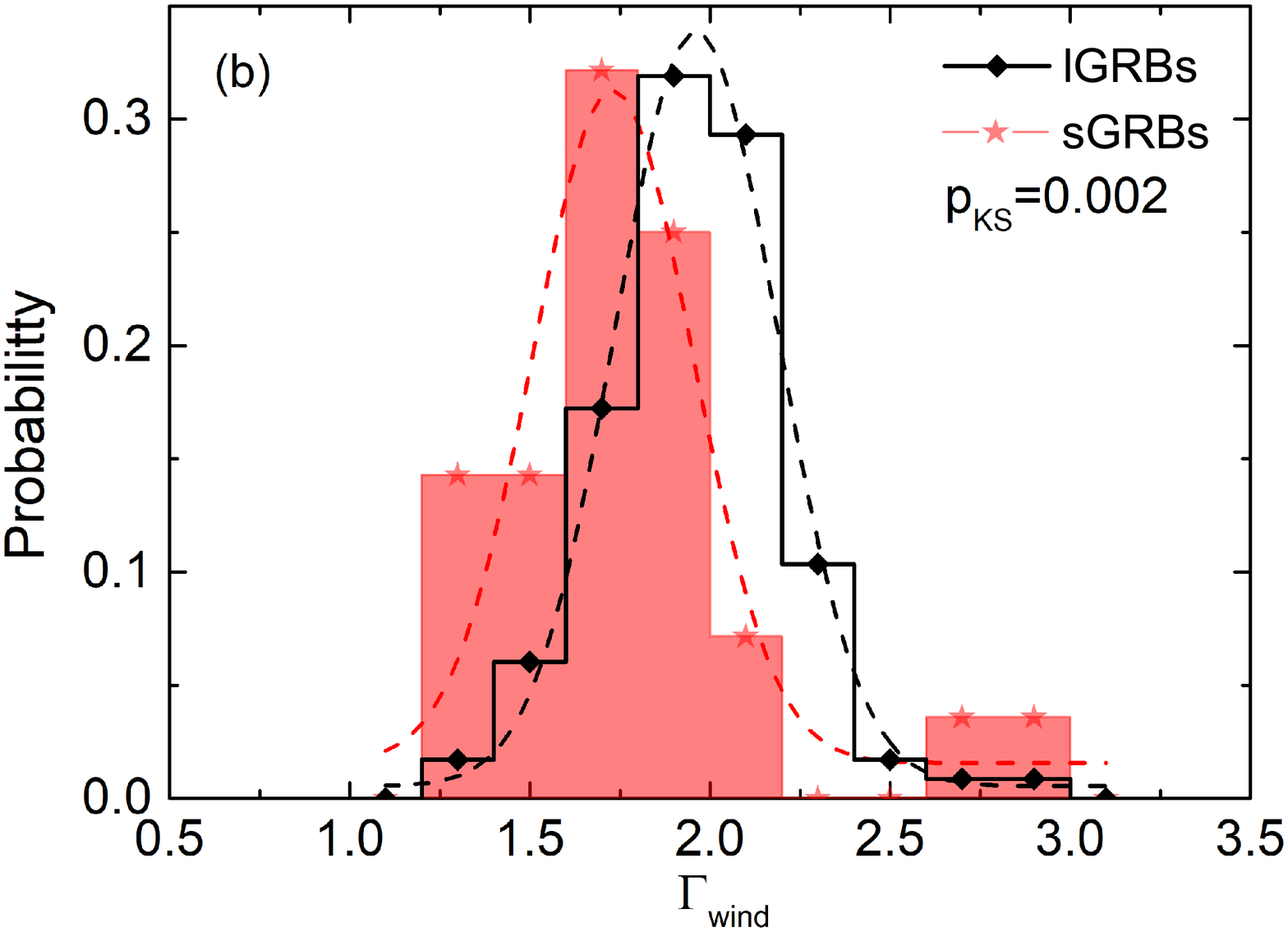}
\center\caption{Comparison of the $\Gamma_{\rm jet}$ and $\Gamma_{\rm wind}$ distributions for the sGRBs and lGRBs. The dashed lines are the best Gaussian fits to the corresponding distributions. The probability of the K-S test for examining the difference of the distribution is also marked.}
\label{gamma}
\end{figure*}

\begin{figure*}
\includegraphics[angle=0,scale=0.5]{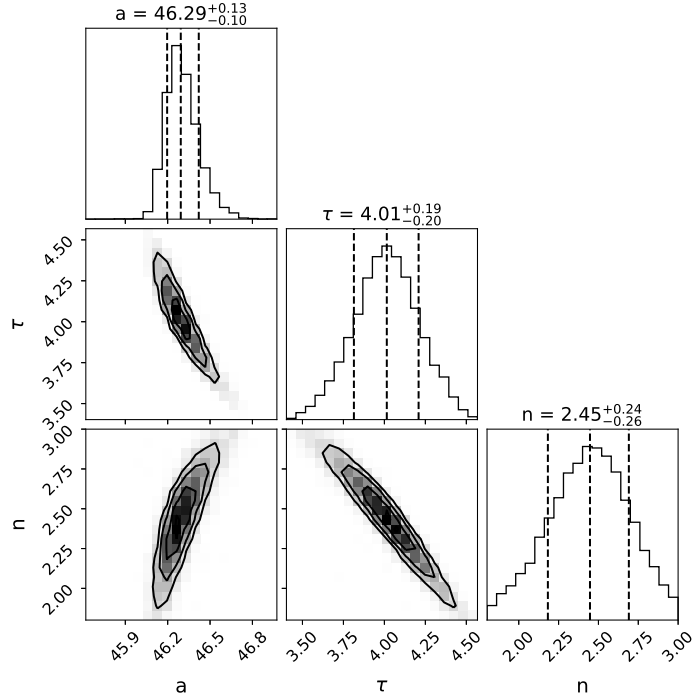}
\center\caption{Probability contours derived from our MCMC fit for deriving the brake index ($n$) and characteristic spin down time scale ($\tau$) of a magnetar in GRB 130603B. The vertical dashed lines mark the $1\sigma$ confidence level of the parameters.}
\label{n}
\end{figure*}

\begin{figure*}
\includegraphics[angle=0,scale=0.35]{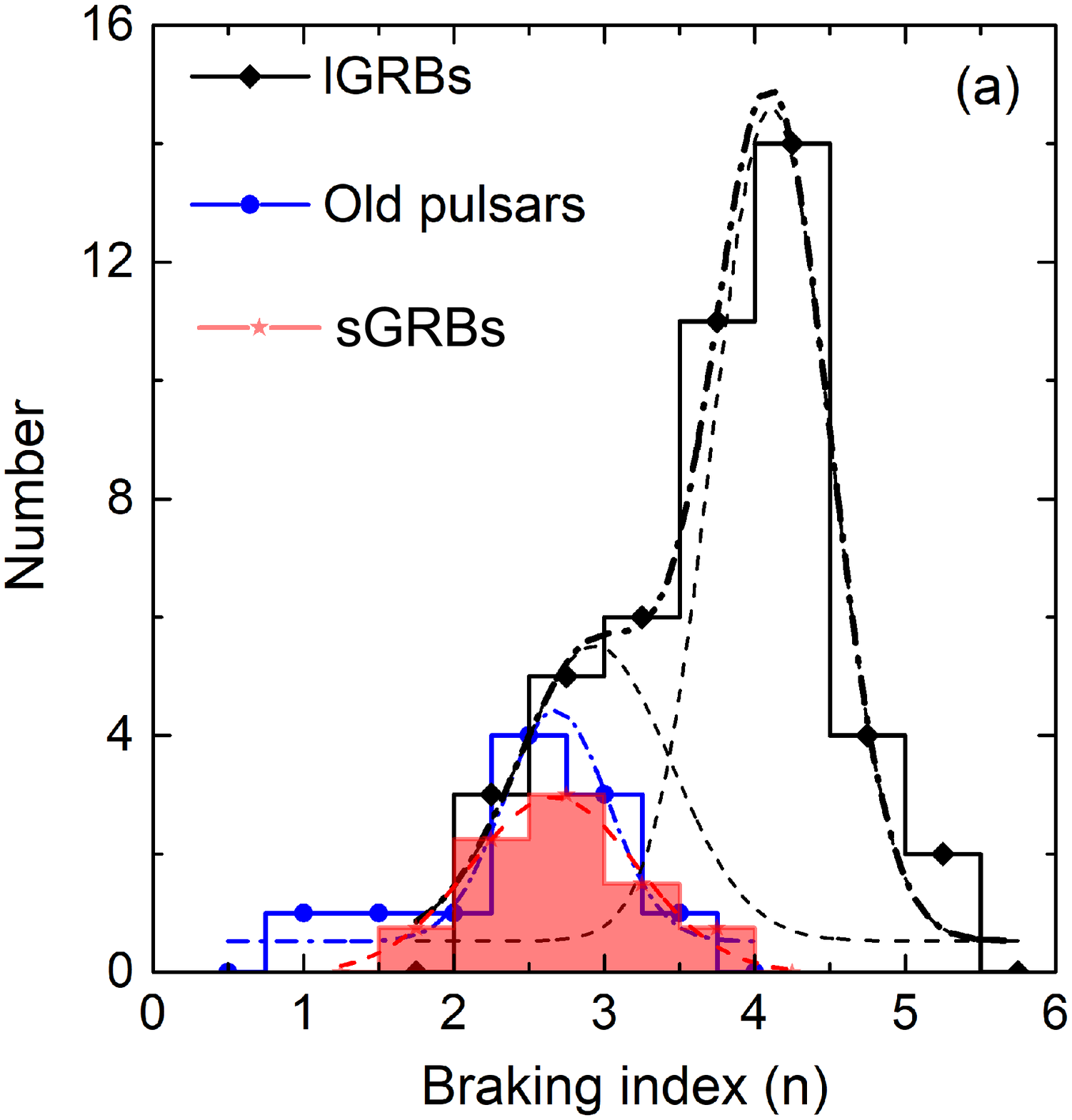}\\
\includegraphics[angle=0,scale=0.35]{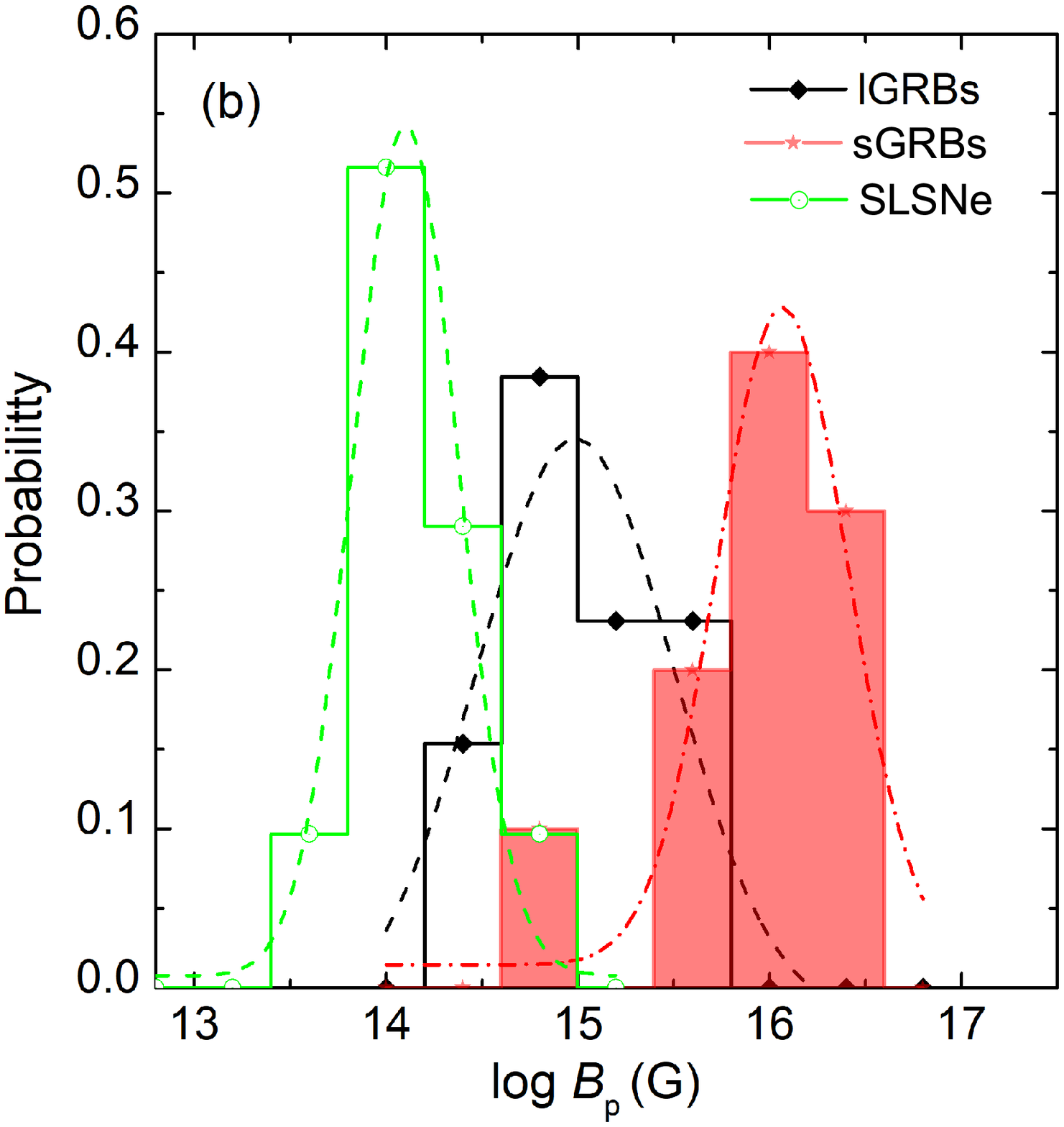}
\includegraphics[angle=0,scale=0.35]{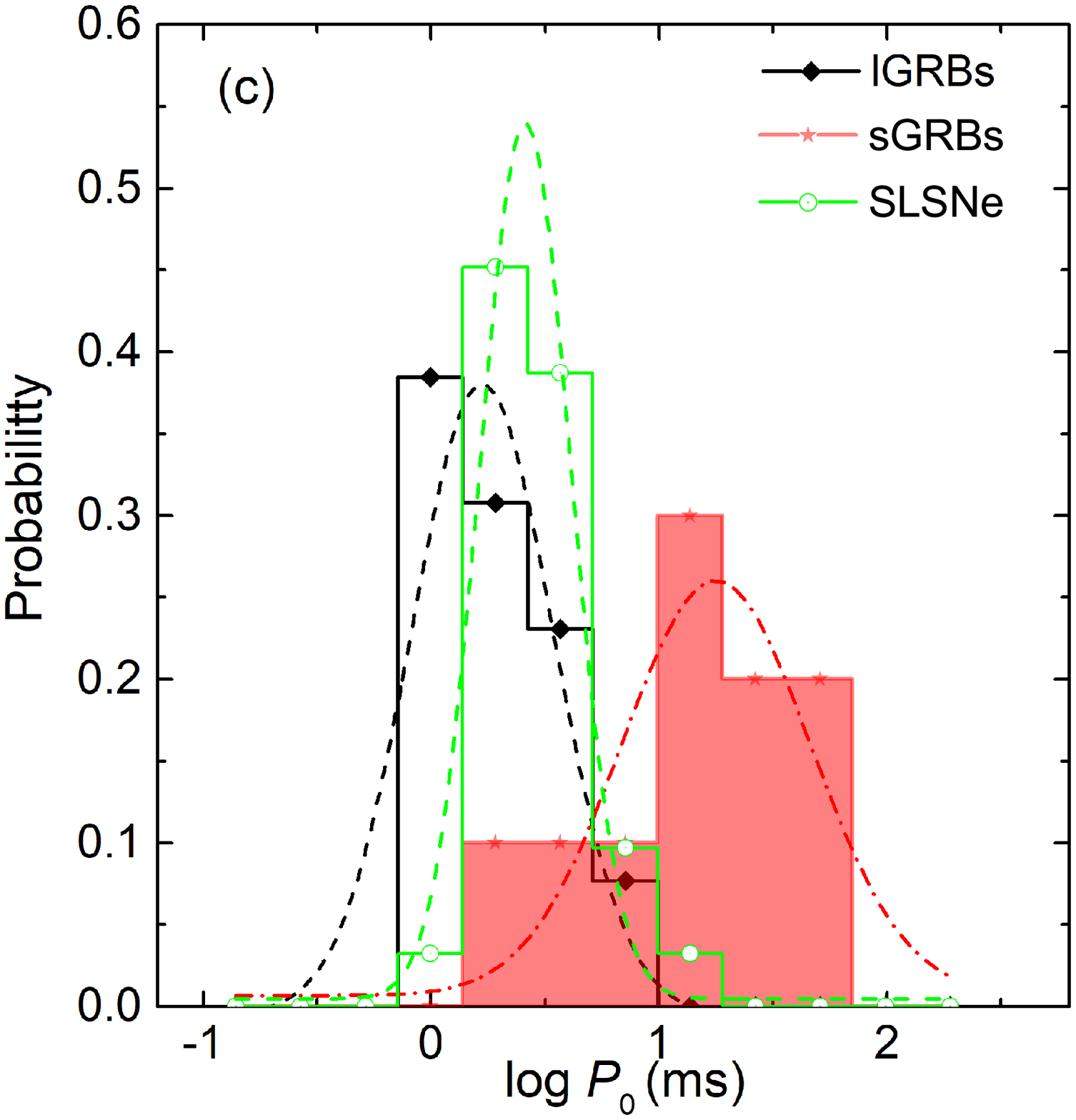}
\center\caption{Panel a---Distribution of the braking index $n$ of the magnetars in the sGRBs in comparison with that for a lGRBs sample from Zou et al. (2019) and a sample of old pulsars from L{\"u} et al. (2019). Panel b and c---Distributions of $B_{p}$ and $P_{0}$ of the magnetars in the sGRBs in comparison with that for a lGRBs sample from Zou et al. (2019) and a sample of superluminous supernovae from Yu et al. (2017).}
\label{derived parameters}
\end{figure*}

\begin{figure*}
\center
\includegraphics[angle=0,scale=0.28]{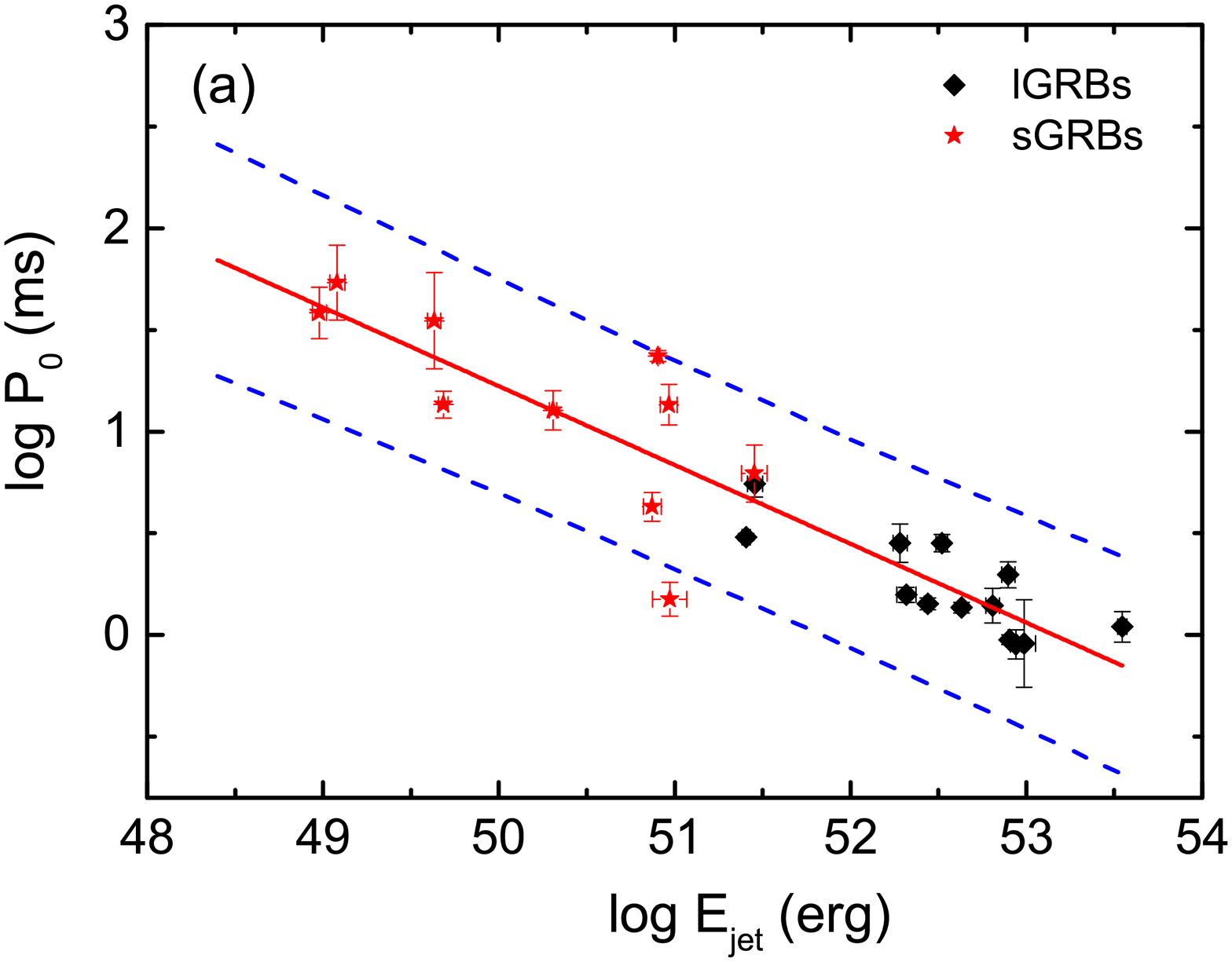}
\includegraphics[angle=0,scale=0.28]{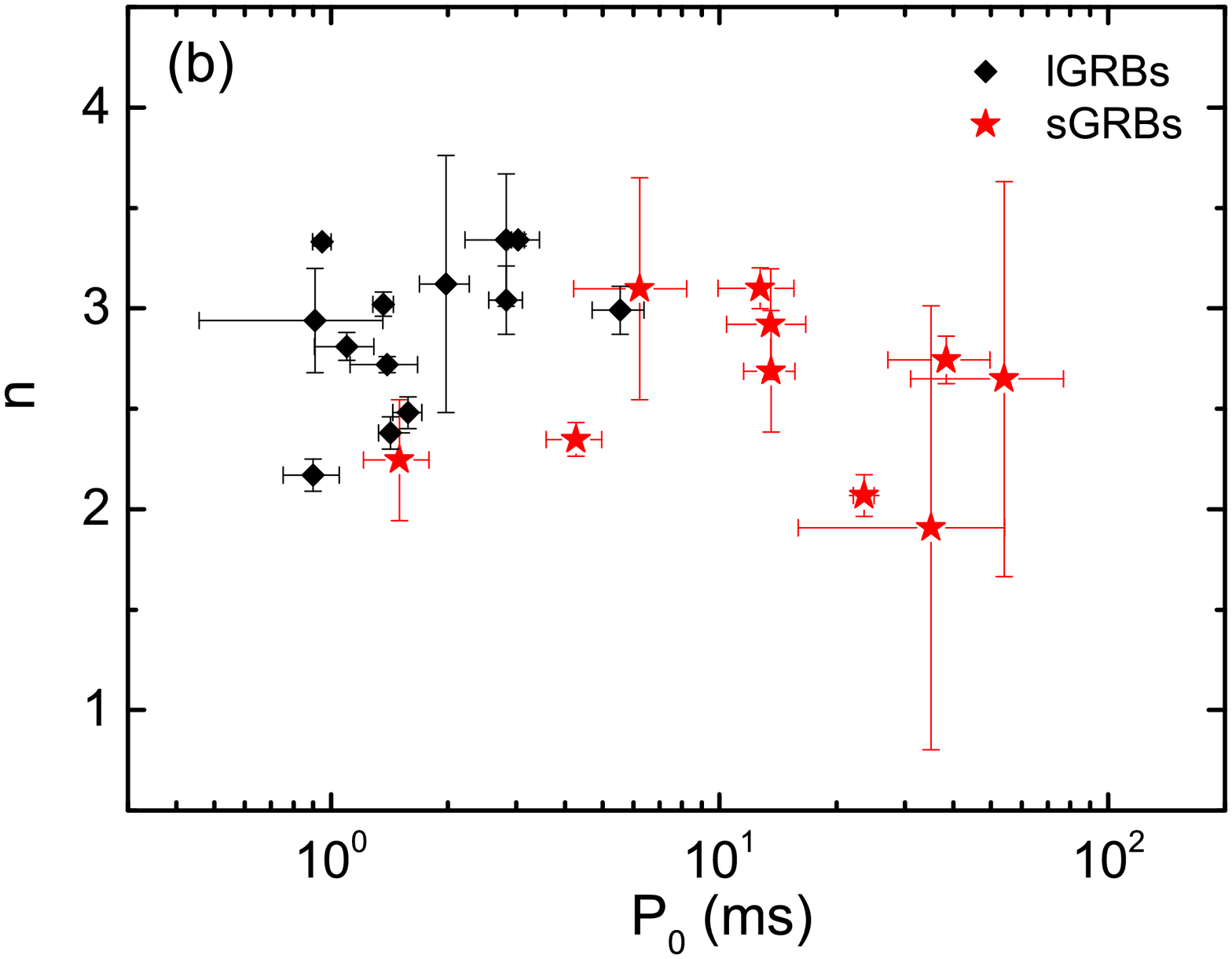}
\caption{$P_0$ as a function of $E_{jet}$ and $n$ for the sGRBs and lGRBs sample. The parameters of lGRBs sample are taken from Zou et al. (2019). The solid and dashed lines are the least-squares linear fits and the 95\% confidence levels for the combined sample of both the sGRBs and lGRBs.}
\label{p-n}
\end{figure*}

\end{document}